\documentclass[pre,aps,twocolumn,preprintnumbers,superscriptaddress]{revtex4}
\usepackage{graphicx}
\usepackage{dcolumn}
\usepackage{bm}
\usepackage{amssymb}
\usepackage{pifont}
\usepackage{amsmath}
\usepackage{times}
\usepackage[nooneline]{subfigure}

\usepackage{graphicx,psfrag}
\usepackage[dvipsnames]{xcolor}
\usepackage{hyperref}
\usepackage[capitalise]{cleveref}

\newcommand\myshade{85}
\colorlet{mylinkcolor}{violet}
\colorlet{mycitecolor}{YellowOrange}
\colorlet{myurlcolor}{Aquamarine}

\hypersetup{
  linkcolor  = mylinkcolor!\myshade!black,
  citecolor  = mycitecolor!\myshade!black,
  urlcolor   = myurlcolor!\myshade!black,
  colorlinks = true,
}

\begin{document}

\title{Low-dimensional model for adaptive networks of spiking neurons}

\author{Bastian Pietras}
\affiliation{Neuronal Dynamics Group, Department of Engineering, Universitat Pompeu Fabra, 08018 Barcelona, Spain.}
\author{Pau Clusella}
\affiliation{EPSEM, Departament de Matemàtiques, Universitat Politècnica de Catalunya, Manresa, Spain.}
\author{Ernest Montbri\'o}
\affiliation{Neuronal Dynamics Group, Department of Engineering, Universitat Pompeu Fabra, 08018 Barcelona, Spain.}

\date{\today}

\begin{abstract}
We investigate a large ensemble of Quadratic Integrate-and-Fire (QIF) neurons with heterogeneous 
input currents and adaptation variables. Our analysis reveals that for a specific class of adaptation, 
termed quadratic spike-frequency adaptation (QSFA), the high-dimensional system can be 
exactly reduced to a low-dimensional system of ordinary differential
equations, which describes the dynamics of three mean-field variables: 
the population's firing rate, the mean membrane potential, and a mean adaptation variable. 
The resulting low-dimensional firing rate equations (FRE) 
uncover a key generic feature of heterogeneous networks with spike frequency 
adaptation: Both the center and the width of the distribution of the neurons' firing frequencies  
are reduced, and this largely promotes the emergence of collective synchronization in the network.   
Our findings are further supported by the bifurcation analysis of the FRE, which accurately captures 
the collective dynamics of the spiking neuron network, including phenomena such as collective oscillations, 
bursting, and macroscopic chaos.
\end{abstract}
\maketitle

\section{Introduction}

Neuronal firing rate equations (FREs) are mathematical descriptions of the collective activity of 
large ensembles of neurons, typically in form of one or a few ordinary differential equations~\cite{WC72,ET10,DJR+08,CW23}. These population models offer an approximate, coarse-grained 
description of the dynamics of spiking neuron networks---generally applicable near asynchronous states---and serve as valuable tools for both theoretical and computational analyses of large-scale neuronal dynamics.

Over the last decade, a singular class of firing rate equations has been obtained~\cite{MPR15,LBS13}. 
These models, often referred to as 'Next-Generation Neural Mass Models' \cite{CB19}, 
are derived exactly from large networks of heterogeneous Quadratic Integrate-and-Fire (QIF) 
neurons, and offer two key advantages over traditional firing rate models: 
First, they provide an exact link between the microscopic dynamics of individual spiking neurons 
and the evolution of two macroscopic variables
---mean firing rate and mean membrane potential. Second, they fully capture both transient 
dynamics and synchronous states in spiking neuron networks. 
Furthermore, the mean-field theory used to derive these exact firing rate equations, which
is closely related to the Ott-Antonsen theory for populations of phase oscillators~\cite{OA08}, 
is versatile enough to accommodate additional biological realism
~\cite{CB19,Lai14,Lai15,RP16,PM16,DRM17,PDR+19,MP20,CKG+23,Pie24,PP22,CM24,PP23,PCP23,GPK23,GVT21,Gol21,VSG+22}.
As a result, these models have become very useful to investigate neuronal dynamics
~\cite{DEG17,BAC19,SA20,ADE22,TAD22,KK22,LO23,BJM+24}, and are powerful modeling tools in neuroscience 
~\cite{BBC17,SAM+18,DG19,KBF+19,SBO+20,GTS+21,RFB+21,RH22,BRN+22,SZM+23,GSK24,NVV24}.

A significant theoretical challenge remains in extending the theory to derive exact FREs 
for populations of QIF neurons with additional dynamic variables
\footnote{See~\cite{CM24i,Paz24} for an interesting line of research in this direction.}.
Several recent studies have developed approximate FREs seeking to describe the 
collective dynamics of such `extended' QIF neurons~\cite{GSK20,FAT+23,CC22,TTO20,GKS21,GVG22,XWQ+23,ADE+24,CC24,DDA+24,SJ24}. 
A particular example are ensembles of QIF neurons with 
spike-frequency adaptation (SFA) \cite{GSK20,FAT+23}, which is a prominent feature of neuronal 
dynamics by which many neuron types reduce their firing frequencies 
in response to sustained current injection, see e.g.~\cite{BH03,GZ14,BT22,Boe17}. 
While an exact mean-field reduction of heterogeneous QIF neurons with SFA remains elusive, 
some studies have approximated the QIF neuron model with SFA 
by assuming that the neuron-specific adaptation variables can be
represented by a global adaptation variable that evolves according to the population's firing rate~\cite{GSK20,FAT+23}. 
This approximation allows for an exact reduction of the spiking neuron network 
to a system of FREs, incorporating the additional adaptation dynamics, 
and captures key collective phenomena that are reminiscent of spiking networks with SFA, 
such as the emergence of collective synchronization (due to the presence of a slow negative feedback), 
and bursting.

However, key aspects of the microscopic dynamics associated with the neuron-specific nature of SFA are not adequately captured by such firing rate models. One overlooked phenomenon arises in populations of neurons with SFA and heterogeneous firing frequencies:
neurons with intrinsically high firing rates undergo a more pronounced reduction in 
firing frequency due to SFA compared to neurons with  lower firing rates. 
As a result, the overall level of frequency heterogeneity diminishes, 
significantly promoting the emergence of collective synchronization~\cite{Win67,Kur84}.

In this work, we take an alternative approach to reduce the dynamics of an extended QIF 
model with a specific form of SFA, termed Quadratic Spike-Frequency Adaptation (QSFA)~\cite{BH03}, 
to an effectively  one-dimensional QIF model 
~\footnote{A similar approach was used by Ott \& Antonsen in Ref.~\cite{OA17},
to investigate a variant of the Kuramoto model with a mechanism of frequency adaptation. 
We emphasize that in the model proposed by Ott \& Antonsen 
the oscillator natural frequencies adapt to the frequency of the
Kuramoto order parameter. 
By contrast, in SFA the degree of adaptation is independent of the level of 
synchrony, and depends only on the firing frequency of each neuron.}.
This allows for analytical progress and the exact 
derivation of a low-dimensional system of FREs for large networks of heterogeneous QIF 
neurons with QSFA. 
In our approach, the adaptation variables remain neuron-specific, 
ensuring that neurons with higher intrinsic firing rates undergo greater adaptation 
than those with lower firing rates. 
This is reflected in an adaptation-induced reduction in the level of heterogeneity 
in the FREs, significantly enhancing the emergence of collective synchronization in the network.

The paper is structured as follows: In \cref{sec:micro}, we introduce the generalized QIF model with 
SFA and describe the approximations leading to the QSFA model. We also illustrate the effects of 
QSFA on the steady states of QIF neuron populations with heterogeneous inputs, demonstrating that 
QSFA results in both a shift and a narrowing of the firing rate distribution. 
In \cref{sec:fre}, we outline the derivation of the FREs for a heterogeneous population of QIF neurons with QSFA. 
In \cref{sec:collective_dyn}, we analyze the bifurcations in the QIF-FRE model with QSFA and
present phase diagrams that summarize the model's possible dynamic regimes. 
Additionally, we compare numerical simulations of the microscopic QIF network with those of the low-dimensional QIF-FRE model. Finally, in \cref{sec:discussion}, we summarize and discuss our findings.

\section{Populations of heterogeneous QIF neurons with spike-frequency adaptation (SFA)}
\label{sec:micro}

We consider a population of  
$N$ neurons with membrane potentials $V_{j=1,\dots, N}$, 
and membrane time constant $\tau_m$,
which evolve according to the following Quadratic Integrate-and-Fire (QIF) model~\cite{EK86,Izh07,LRN+00}
\begin{subequations}
\label{model0}
\begin{eqnarray}
\tau_m \dot V_j &=& V_j^2 + I_j - a_j , \label{model0V}\\
\tau_a \dot a_j &=& -a_j +\beta  f_j.	
\label{model0a}
\end{eqnarray}
\end{subequations}
The last two terms on the r.h.s.\ of Eq.~\eqref{model0V} vary from neuron to neuron and 
represent, respectively, constant inputs and adaptation currents of strength $\beta\geq 0$.    
The definition of the QIF model requires a resetting rule such that after each spike 
---which is marked by the spike time $t_j^k$ at which $V_j$ reaches infinity---, the voltage 
is instantaneously reset to minus infinity. For the spike resetting at infinity, 
the spike-frequency (or firing rate) of intrinsically active neurons
($I_j - a_j>0$)  is~\cite{EK86,Izh07}
$$\nu_j=\tfrac{1}{\pi\tau_m}\sqrt{I_j -a_j},$$
and $\nu_j=0$ for quiescent, or excitable, neurons ($I_j- a_j\leq 0$).
The adaptation variables $a_j$ obey the linear, 
first order differential equations Eq.~\eqref{model0a}, 
where $f_j$ measures the frequency of the spikes of neuron $j$.
Spike-frequency adaptation (SFA) is often modeled by
substituting the term $f_j$ in Eq.~\eqref{model0a} with the spike train of neuron $j$, 
so that the adaptation variable $a_j$ increases by a finite amount $\beta/\tau_a$ whenever neuron $j$
undergoes an action potential
~\cite{BH03,GZ14,BT22,Boe17,Wan98,FMT02,LRN+00,GMG07,SL13,ALB+17,SDG17,dVRC+19,VH01}; 
if neuron $j$ does not spike, $a_j$ decays to zero with 
time constant $\tau_a \gg \tau_m$. 

An important dynamical consequence of spike-dependent adaptation models is that they 
only slow down the firing frequency of intrinsically firing neurons 
($I_j-a_j>0$), but cannot stop their repetitive firing~\cite{GZ14};
certainly, spike-dependent adaptation cannot initiate firing in 
those neurons that are intrinsically quiescent ($I_j-a_j\leq0$), either. 
Hence, while the number of firing neurons remains the same,
this dynamical feature changes 
the distribution of the neurons' firing frequencies by reducing not 
only its mean, but also its \emph{width}.

\subsection{Quadratic spike-frequency adaptation (QSFA) model}

%
\begin{figure}[t]
\centering
\includegraphics[width=\columnwidth]{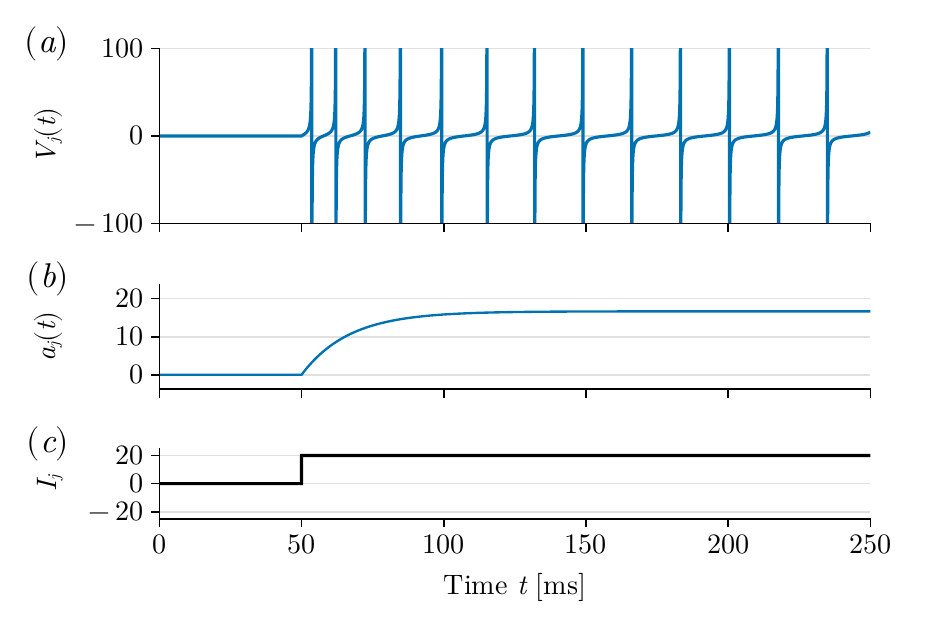}
\caption{ Dynamics of a QIF neuron with QSFA, Eqs.~(\ref{model0}\!~\&~\!\ref{f}). 
A spike train of an adapting neuron, in panel (a), is evoked by 
the onset of the stimulus shown in panel (c). 
Panel (b): For $t\geq 50$, the adaptation variable evolves as: 
$a_j(t)= I_j\beta/(1+\beta)   (1-e^{-(t-50) (1+\beta)/\tau_a})$,
and converges to $a_j^*=50/3$, see \cref{aFP}.
As adaptation builds up, the frequency of the spikes drops from an initially 
high onset rate to a lower, steady-state frequency given by \cref{fI}. 
Parameters:
$\beta=5$ and $\tau_a = 10\tau_m=100$ms. 
}
\label{fig:single_neuron}
\end{figure}

To simplify the analysis of the QIF model Eqs.~\eqref{model0}, one may replace
the discontinuous spike train of the spike-dependent adaptation 
model by a continuous, linear function of the instantaneous spike-frequency
$\nu_j$, i.e.~$f_j \propto \nu_j$, see, e.g.~\cite{Erm98lin}. 
Alternatively, here we propose the following \emph{quadratic} 
spike-frequency adaptation (QSFA) model, 
\begin{equation}
f_j=  I_j - a_j,
\label{f}
\end{equation}
in which $f_j$ is proportional to the square of the spike-frequency of 
those neurons that are intrinsically active, i.e.~$f_j \propto \nu_j^2$. 
\cref{fig:single_neuron}(a) shows the time series of the voltage variable $V_j$ of a 
quiescent QIF neuron with QSFA that receives a step input current at $t=50$ms and becomes self-oscillatory.
Initially, the adaptation variable is $a_j(0)=0$, and then exponentially converges to 
the fixed point of Eq.~\eqref{model0a},  
\begin{equation}
a_j^*=\tfrac{\beta }{1+\beta} I_j.
\label{aFP}
\end{equation}
Accordingly, the steady-state frequency the QIF neuron 
(often referred to as the neuron's f-I curve) is
\footnote{The effect os QSFA is 
the rescaling of the f-I curve by a factor $1/\sqrt{1+\beta}<1$. The square root input dependence 
of Eq.~\eqref{fI} is characteristic of quadratic SFA models~\cite{BH03}, 
whereas in linear SFA models (with $f_j\propto \nu_j$), 
the f-I curve scales linearly with the input near the onset of 
firing~\cite{Erm98lin,Wan98}.}
\begin{equation}
\nu_j=\tfrac{1}{\pi\tau_m} \sqrt{\tfrac{I_j }{1+\beta}},
\label{fI}
\end{equation}
if $I_j >0 $, and $\nu_j=0$ otherwise. 
\cref{fI} shows that it is exclusively the sign of $I_j$ that determines the dynamical 
character of each neuron: QSFA either slows down the firing rate 
of intrinsically active neurons ($I_j>0$) without stopping firing, or it brings quiescent neurons ($I_j<0$)
closer to their firing threshold, yet without inducing firing. The ratio between active and quiescent neurons thus remains the same. 
And while the frequencies $f_j$ in \cref{f} become negative for $I_j<0$,
this only alters the shape of the inputs' \emph{subthreshold} distribution, but does not influence the level of activity of the population.

The choice of the QSFA model \eqref{f} has two benefits 
that critically simplify the study of the mean-field population model: 
First, Eqs.~\eqref{model0a} become 
independent of the particular state of neuron $j$, so that the dynamics of the 
QIF neurons Eqs.~\eqref{model0V} becomes effectively one-dimensional.
Second, due to the quadratic dependence of $f_j$ on the neuron's frequency, 
the adaptation variables acquire the same distribution type as that of the parameters 
$I_j$. 
In particular, we will show that if both $I_j$ and $a_j(0)$ 
are distributed according to a Lorentzian
distribution, the variables  $a_j$ remain Lorentzian distributed at all times.
Notably, this allows us to apply the technique originally proposed in~\cite{MPR15} to derive an 
exact, low-dimensional system of FREs which exactly describes the dynamics of a population of 
QIF neurons with QSFA Eqs.~(\ref{model0}\!~\&\!~\ref{f}) in the $N\to \infty$ limit. 

\subsection{Effect of QSFA on the distribution of firing frequencies}
\label{sec:SFAstatistics}

\begin{figure}[t]
\centering
\includegraphics[width=\columnwidth]{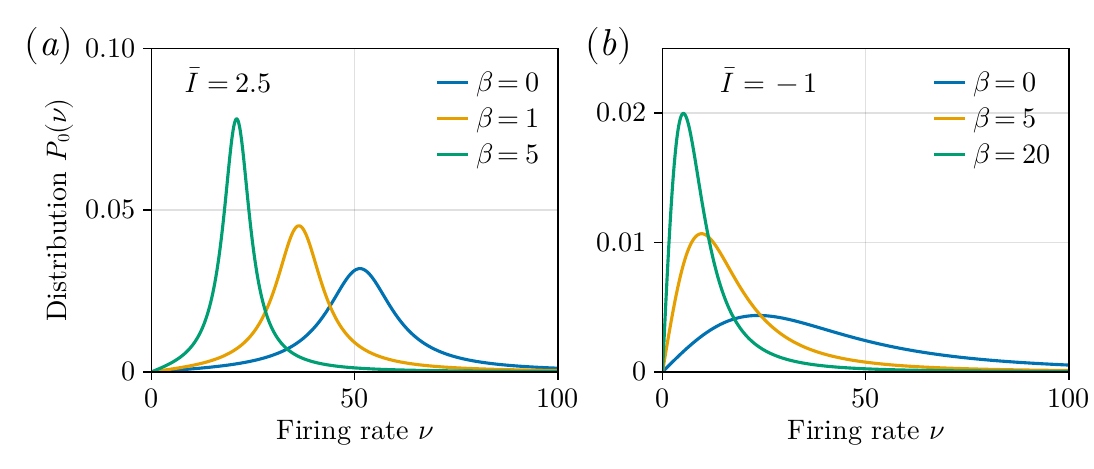}
\caption{Quadratic Spike-Frequency Adaptation (QSFA) reduces both the center and the width of the firing frequency distributions in populations of heterogeneous QIF neurons.
We show the graph of Eq.~\eqref{eq:Peta} with a Lorentzian distribution of currents 
$g(I)$, for different levels of the adaptation strength $\beta$, and for 
a population where the majority of neurons are: 
(a) intrinsically spiking neurons, $\bar I=2.5$;
(b) quiescent neurons, $\bar I=-1$. 
In each case, the area under the three graphs is the same, indicating that QSFA 
does not alter the proportion of intrinsically spiking neurons in the population.
Parameters: $\Delta=1$, $\tau_m=10$ms.}
\label{fig:fr_distributions}
\end{figure}

Before starting the derivation and analysis of the FRE with QSFA, it is 
illustrative to investigate the effect of QSFA on the 
steady-state distribution of firing frequencies of a population of 
(non-interacting) QIF neurons.
We begin by identifying two important outcomes of Eq.~\eqref{fI} 
that generally occur in populations of extended QIF neurons Eqs.~\eqref{model0}: 
Both the center and the width of the firing frequency distribution asymptotically 
shift to zero as the level of SFA is increased. 
That is, an overall decrease of activity in the population 
is accompanied by a global homogenization of the firing rates, compensating 
for the population's intrinsic heterogeneity.   
For the special case of QSFA, Eq.~\eqref{aFP} shows the important property that 
the fixed point values $a^*_j$ acquire the same distribution 
type as that of parameters $I_j$, where both the center and the width of the $a^*_j$ distribution are scaled by the factor $\beta/(1+\beta)$.
Effectively, this leads to a rescaling of both the center and the width of the $I_j$ distribution by $1/(1+\beta)$.
In consequence, 
the proportion of firing (or quiescent) neurons in the population 
is determined solely by the distribution of inputs $I_j$.

Finally, we explicitly compute the firing frequency distribution for QIF neurons with QSFA. 
Given a distribution $g(I)$ of inputs $I_j$, 
the (stationary) distribution of firing rates $P_0(\nu) = g(I) | dI/d\nu|$, with 
$I(\nu) = (1+\beta) (\pi\tau_m \nu)^2$, satisfies
\begin{equation}
P_0(\nu)=2 (1+\beta) (\pi\tau_m)^2 \nu \, g\big( (1+\beta) (\pi\tau_m\nu)^2 \big).
\label{eq:Peta}
\end{equation}
In \cref{fig:fr_distributions}, we show how this distribution 
changes with increasing 
adaptation strength $\beta>0$ for a Lorentzian distribution of inputs $I_j$, 
$g(I)=\Delta/\pi[(I-\bar I)^2+\Delta^2]^{-1}$, of width $\Delta$, and centered at positive ($\bar I=2.5$) and negative 
($\bar I=-1$) values. Increasing $\beta$ 
shifts the center of the distribution to the left, and reduces its width. 
Integration of $P_0(\nu)$ shows that the area under the graphs is 
independent of $\beta$. This indicates that QSFA does not alter the proportion of 
intrinsically spiking neurons 
in the population, which is solely determined by $g(I)$
\footnote{
The integral 
$$
\int_0^\infty P_0(\nu) d\nu=\frac{1}{2} +\frac{1}{\pi}\arctan\left(\frac{\bar I}{\Delta}\right),
$$
gives the proportion of intrinsically spiking neurons in the ensemble, which   
solely depends on the parameters of the distribution of heterogeneity, $g(I)$ ---and not on $\beta$.}

\section{Firing Rate Equations With Quadratic Spike Frequency Adaptation}
\label{sec:fre}

In \cref{sec:micro}, we have shown that QSFA strongly shapes the distribution of spike frequencies   
in populations of QIF neurons with distributed inputs.
In the following, we demonstrate that this greatly influences the synchronization properties 
of large networks of recurrently coupled spiking neurons.  

To investigate non-trivial collective dynamics of the QIF network Eqs.~(\ref{model0}\!~\&\!~\ref{f}), 
we first extend our model so that neurons 
are able to interact with each other via a mean-field  coupling. 
Specifically, hereafter we investigate the model Eqs.~\eqref{model0} with
\begin{equation}
I_j (t) =   J \tau_m R(t) + \eta_j. 
\label{I}
\end{equation}
The first term consists of a mean-field excitatory coupling of strength $J>0$. 
This coupling term is mediated by the population firing rate $R(t)$, 
which is obtained from the spike count function 
$$
S(t) = \frac1N \sum_{j=1}^N \sum_k \frac1{\tau} \int_{t-\tau}^t \delta(s-t^k_j) ds,
$$
as $\lim_{\tau\to0} S(t)=R(t)$.  
The terms $\eta_j$ 
represent constant inputs 
that vary from neuron to neuron according to a Lorentzian distribution centered 
at $\bar\eta$ with half-width at half-maximum $\Delta >0$:
\begin{equation}
g(\eta)= \frac{\Delta/\pi}{(\eta-\bar \eta)^2+\Delta^2}.
\label{lor}
\end{equation}

We now derive a low-dimensional system of differential equations 
(the so-called FREs) governing the 
evolution of the population firing rate $R$ and mean voltage $V$ of 
the population of QIF neurons. 
To this end, we first decompose the general 
solution of the linear ordinary differential equation 
Eq.~\eqref{model0a} with QSFA \eqref{f}
\begin{equation}
\tau_a \dot a_j  = -(1+\beta) a_j + \beta \left[ J\tau_m R(t)+ \eta_j\right],
\label{a_ODE}
\end{equation}
into two parts, as  
\begin{equation}
a_j(t)=c_j e^{-t/\tau}+\alpha_j(t).
\label{a_sol}
\end{equation}
Since the first part of the solution decays exponentially to zero, 
the specific choice of the constants of integration $c_j$ is irrelevant after a 
transitory period of the order of the lifetime
\begin{equation}
\tau = \frac{\tau_a}{1+\beta}.
\label{tau}
\end{equation}
Still, for reasons that will become clear shortly, 
hereafter we consider that $c_j$ are distributed according to a Lorentzian distribution
centered at $\bar c$ with half-width at half-maximum $\gamma >0$ 
\begin{equation}
f(c)=\frac{\gamma/\pi}{(c-\bar c)^2+\gamma^2}.
\label{lorentzian2}
\end{equation}
The second component of the solution Eq.~\eqref{a_sol}, $\alpha_j(t)$, is the 
particular solution of Eq.~\eqref{a_ODE} with $a_j(0)=0$.
It is important to note that for Lorentzian distributed inputs $\eta_j$, the adaptation 
variables $\alpha_j(t)$ are also Lorentzian distributed
~\footnote{
The particular solution of the linear differential 
Eq.~\eqref{a_ODE} with $a_j(0)=0$ has the form 
\begin{equation}
\alpha_j(t)= \eta_j h(t) + l(t) ,
\label{Aj_int}
\end{equation}
with  
$l(t) =\beta J \tau_m/\tau_a \int_0^t e^{(t'-t)/\tau}  R(t') dt',$
and $h(t)= (1 -  e^{-t/\tau})\beta/(1+\beta)$.
Thus, variables $\alpha_j(t)$ have the same distribution type as that of parameters 
$\eta_j$.   
}.
Substituting Eqs.~(\ref{I},\ref{a_sol}) into Eq.~\eqref{model0V} yields the QIF model   
\begin{equation}
\tau_m \dot V_j= V_j^2 +  J \tau_m R(t) + \eta_j - c_j e^{-t/\tau}- \alpha_j(t),
\label{model0Vi}
\end{equation}
where $\eta_j$, $c_j$, and $\alpha_j(t)$ 
are all distributed according to Lorentzian probability density functions.

Equation~\eqref{model0Vi} belongs to a class of mean field models that in the limit $N\to \infty$  
admit an exact, low-dimensional description in terms of the population mean 
firing rate and membrane potential ---see Eq.~(19) in Ref.~\cite{MPR15}. 
In the following we derive such low-dimensional FRE using the procedure originally proposed 
in~\cite{MPR15}. Accordingly, we adopt the thermodynamic 
limit of Eqs.~\eqref{model0} and drop the indices in Eqs.~(\ref{a_ODE},\ref{model0Vi}).
We denote by $\rho(V|\eta,c,t)$ the density of neurons with voltage $V$, 
given parameters $\eta$ and $c$, whose evolution is governed by the continuity equation
\begin{equation}
\tau_m \partial_t \rho + \partial_V \left[\rho  \left( V^2+ J \tau_m R+ \eta 
-   c~ e^{-t/\tau} - \alpha  \right)  \right]=0.
\label{cont}
\end{equation}
Substituting the `Lorentzian ansatz' 
\begin{equation}
\rho(V \vert \eta,c, t)= \frac{1}{\pi}\frac{X(\eta,c,t)}{\left[V-Y(\eta,c,t)\right]^2 +X(\eta,c,t)^2}
\label{la}
\end{equation}
into \cref{cont}, we find that, for each value of $\eta$ and $c$,
the variables $X$ and $Y$ satisfy  
\begin{equation}
\tau_m \partial_t W= i\left[  J \tau_m R +\eta - c~ e^{-t/\tau} - \alpha  -W^2\right] , 
\label{w}
\end{equation}
where $W(\eta,c,t)\equiv X(\eta,c,t) +i Y(\eta,c,t)$. 
The population firing rate is related with the variable $X(\eta,c,t)$ as
\begin{equation}
R(t)=\frac{1}{\pi\tau_m}\int_{-\infty}^{\infty} f(c)\int_{-\infty}^{\infty} 
X(\eta,c,t) \;\! g(\eta) \;\!d\eta\;\! dc ,
\label{r}
\end{equation}
and, since $Y(\eta,c,t)$ is the center of the distribution of membrane potentials $\rho(V|\eta,c,t)$,
the (Cauchy principal value of the) integral of $Y$ is the mean membrane potential 
\begin{equation}
V(t)= \int_{-\infty}^{\infty} f(c) \int_{-\infty}^{\infty} Y(\eta,c,t) \;\! g(\eta) \;\!d\eta\;\! dc.
\label{v}
\end{equation}
Eqs.~\eqref{r} and \eqref{v} couple the infinite set of Eqs.~\eqref{w}.   
By considering the analytic continuation of $W$ in the complex $\eta$ and $c$ planes, we 
require $\mathrm{Re}(W)$ to not become negative. 
We thus consider the poles of $g(\eta)$ and $f(c)$ such that $\partial_t \mathrm{Re} 
(W) |_{X =0} > 0$, i.e.~$\eta=\bar\eta-i\Delta$ and $c=\bar c + i\gamma$
\footnote{
This can be seen by substituting \cref{Aj_int} in \cref{w} with $X(t) = R(t)=0$, 
which yields
$$
\tau_m\partial_t W|_{X=0} = i \left[ \eta - \eta h(t) - l(t)  - c e^{-t/\tau} + Y^2 \right].
$$
Then, for $X= \mathrm{Re}(W)$, the derivative $\partial_tX(t)$ evaluated at $X(t)=0$ can be 
evaluated using complex-valued $\eta$ and $c$, as
\begin{align*}
\tau_m\partial_t X|_{X=0} = -\eta_i (1-h) + c_i e^{-t/\tau} \stackrel{!}{>} 0.
\end{align*}
where the subscript $i$ indicates the imaginary parts of $\eta$ and $c$. 
Given that $h\in[0,1]$, this 
can be independently satisfied (either for $\eta_i\neq 0=c_i$, or for $c_i\neq0=\eta_i$) 
only if $\eta_i < 0$ or $c_i > 0$.
}.
Then, by applying Cauchy's residue theorem, we find that
\begin{equation}
W(\bar \eta- i\Delta, \bar c +i \gamma,t) =\pi\tau_m R(t)+i V(t) .
\label{W_pole}
\end{equation}
The dynamics of $R$ and $V$ can be obtained from \cref{w} after expanding the adaptation variable 
$\alpha(\eta,t)$ to the complex $\eta$-plane and evaluating it at the pole of $g(\eta)$, $\eta=\bar\eta-i\Delta$. 
Defining $A$ and $B$ as the real and imaginary parts of $\alpha(t,\bar\eta-i\Delta)$,
\begin{equation}
\alpha(t,\eta)= \alpha(t, \bar \eta- i\Delta) \equiv A(t)+i B(t),
\label{Alpha_pole}
\end{equation}
and substituting 
Eqs.~(\ref{W_pole},\ref{Alpha_pole}) into \cref{w}, yields the firing rate equations 
\begin{subequations}
\label{fre}
\begin{align}
\tau_m \dot R =& \frac{1}{\pi\tau_m} \left[\Delta +   
\gamma e^{-t/\tau} +  B \right] + 2  R V, 
\label{frea}\\ 
\tau_m \dot V =&   V^2 - (\tau_m  \pi R)^2 + \bar \eta + J \tau_m R -  A
-\bar c ~e^{-t/\tau},
\label{freb} 
\end{align}
\end{subequations}
where the initial conditions $R(0) = x_0 / (\pi \tau_m) \geq 0$ and $V(0) = y_0 \in \mathbb R$ 
are related to the width $x_0$ and center $y_0$ of the Lorentzian distribution of 
initial voltage variables $V_j(0)$.
The evolution of the adaptation variable $\alpha$ can be determined by substituting \cref{a_sol} into 
\cref{a_ODE} and then using \cref{Alpha_pole}.
The solution of the imaginary part of the  resulting equation is  
\begin{equation}
B(t) =  \tfrac{\Delta\beta(e^{-t/\tau}-1)}{1+\beta},
\label{Bsol}
\end{equation}
whereas $A(t)$ obeys 
\begin{equation}
\tau_a \dot A = -A (1+\beta)+ \beta \left[\bar \eta+ J\tau_m R(t)   \right],
\label{A_FRE}
\end{equation}
with $A(0)=0$. Then, after a transitory period of the order of $\tau$,
the dynamics of~\cref{fre} converges to the system of FRE
\begin{subequations}
\label{fre3}
\begin{align}
\tau_m \dot R &= \frac{1}{\pi\tau_m}  \frac{\Delta}{1+\beta} + 2  R V, 
\label{fre3a}\\ 
\tau_m\dot V &=   V^2 - (\tau_m  \pi R)^2 + \bar \eta + J \tau_m R   -  A.\label{fre3b}
\end{align}
\end{subequations}
The three-dimensional system Eqs.~(\ref{fre3},\ref{A_FRE}) governs the asymptotic collective 
dynamics of the population of QIF neurons Eqs.~(\ref{model0},\ref{f},\ref{I}), 
where $A(t)$ corresponds to the mean of the adaptation variables $a_j(t)$ 
\footnote{ 
After a transitory period of time $\tau$, the variables $A(t)$ and $|{B}(t)|$ correspond 
 to the mean and the spread of the neurons' adaptation variables $a_j(t)$.
The solution \cref{Bsol} indicates that the width of the distribution of
adaptation variables is reduced by the factor $\beta/(1+\beta)$. 
If $\bar \eta$ is time-independent, the solution of \cref{A_FRE} can be written as 
$A(t)=\bar \eta (1-e^{-t\tau})/(1+\beta) +\tilde A(t)$
where $\tilde A(t)$ follows the dynamics 
$\tau_a \dot {\tilde A} = -\tilde A (1+\beta)+ \beta  J\tau_m R(t)  $.
Therefore, after a transient of the order $\tau$, 
the solution is $A(t) = \bar \eta/(1+\beta) + \tilde A(t)$, and 
the center of the distribution of 
adaptation variables is also reduced by a factor $\beta/(1+\beta)$.
}.

\section{Collective dynamics of populations of QIF neurons with QSFA}
\label{sec:collective_dyn}

\begin{figure}[t]
\includegraphics[width=\columnwidth]{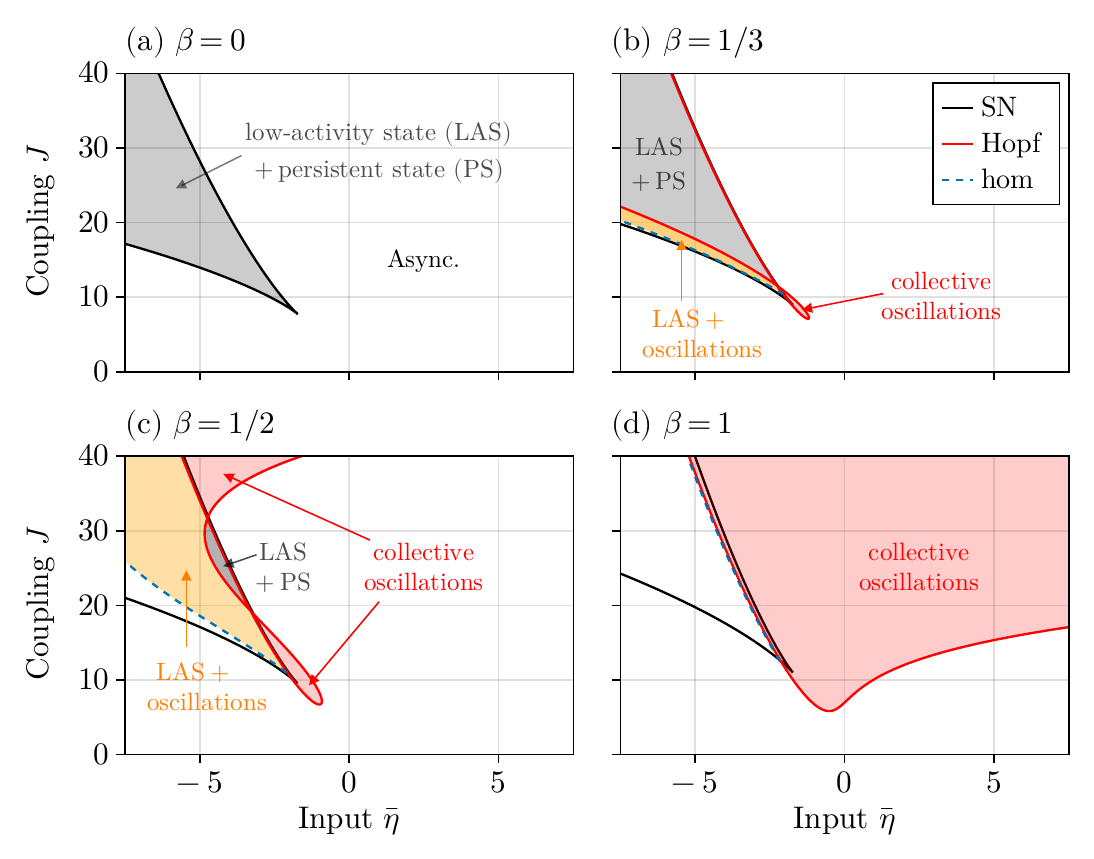}
\caption{Phase diagrams $(\bar\eta,J)$ for increasing values of adaptation strength $\beta$.
The Saddle-Node boundaries (SN, solid black lines) describe a cusp-shaped 
region of coexistence between low activity 
(LAS) and persistent (PS) states. Hopf and Homoclinic bifurcations correspond to the 
red and dashed blue boundaries, respectively.  
White regions: A fixed point corresponding to an asynchronous state 
is the only stable state.
Gray-shaded regions: Bistability between two asynchronous states, LAS and PS. 
Red-shaded regions: Collective oscillations are the only stable state. 
Yellow-shaded regions: Bistability between LAS and collective oscillations. 
See also \cref{fig:largeG}.
Parameters:   $\Delta=1$, $\tau_a=10\tau_m=100$ms.
}
\label{fig:Hopf_bound}
\end{figure}
\begin{figure}[t]
\centering
\includegraphics[width=1\columnwidth]{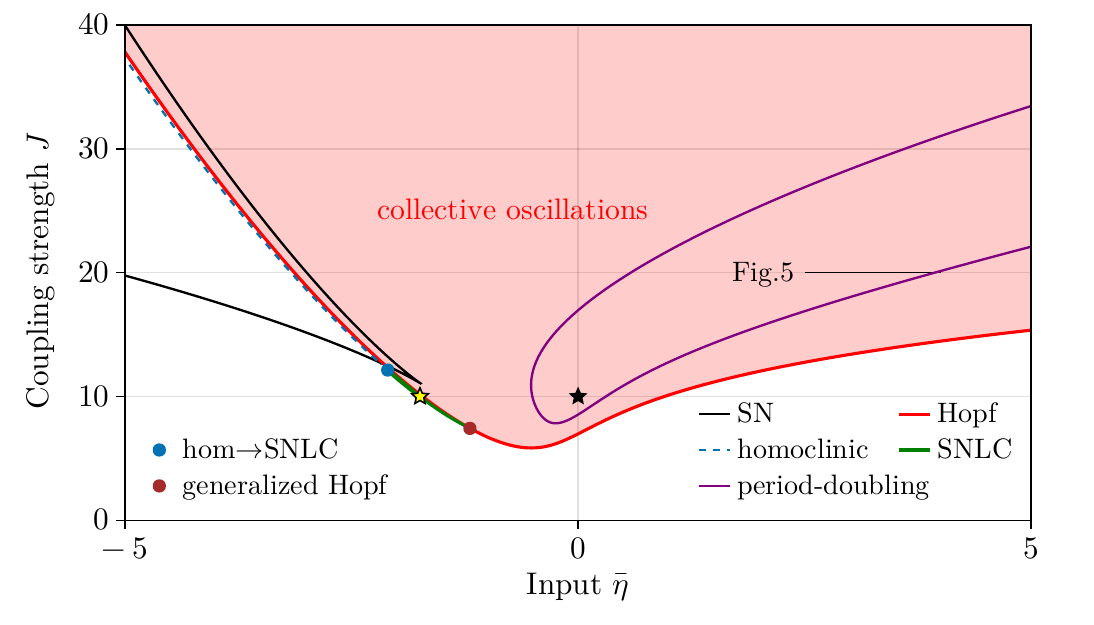}\\
\caption{Enlarged view of the phase diagram in \cref{fig:Hopf_bound}d, 
corresponding to strong adaptation, $\beta=1$.  
The diagram is dominated by the red-shaded region, 
where collective oscillations are the only stable attractor. 
For $\bar\eta<-1$, the Hopf bifurcation (Red lines) becomes subcritical 
at a generalized Hopf point (dark red dot), from where a saddle-node of limit
cycle (SNLC) bifurcation emerges (green line).
Close to the cusp-shaped SN bifurcation lines (black lines), the SNLC curve 
becomes a homoclinic bifurcation (blue dashed line) at the blue dot.
Between the SNLC/homoclinic and the Hopf bifurcation curve, there is bistability 
between LAS and collective oscillations (the yellow star denotes
 the parameters of the numerical simulations shown in \cref{fig:net_sim}).
Collective oscillations emerging from the Hopf curve can undergo secondary bifurcations:
we found a period-doubling bifurcation (purple line, the black star denotes the 
parameters of the numerical simulations shown in \cref{fig:net_sim}) and
within the period-doubling curve, there is a transition to chaotic collective
 dynamics through a period-doubling cascade (see \cref{fig:chaos}).
}
\label{fig:largeG}
\end{figure}

We next analyze the FREs~(\ref{fre3},\ref{A_FRE}) for globally coupled, excitatory QIF 
neurons with QSFA. We focus on the analysis of persistent states (PS), the onset of collective 
oscillations, as well as in the presence of network bursts
---that have been also found in 
spiking neuron networks with alternative models of SFA~\cite{LRN+00,VH01,GMG07,DNC+12,GSK20,FAT+23}---
and collective chaos.

It is well known that strong enough levels of recurrent excitation $J$ 
may generally produce high activity, asynchronous states 
in neural networks---so-called persistent states (PS). 
Although PS may be encountered in the presence of adaptation, we find that they are 
easily destabilized giving rise to oscillatory behavior.
To investigate these instabilities, we first evaluate 
the fixed points of the FREs~(\ref{fre3},\ref{A_FRE}), which we write as~\cite{DRM17}
\begin{equation}
R^*= \Phi \left(  \bar \eta +J \tau_m R^*\right),
\label{Rss}
\end{equation}
where the population's f-I curve is
\footnote{
Similar to the f-I curve of an individual QIF neuron with QSFA, Eq.~\eqref{fI}, 
also the population f-I curve $\Phi$ is a non-negative, monotonously increasing function 
that scales as the square root of the input for large $I$.
}
\begin{equation}
\Phi(I)= \frac{1}{\sqrt{1+\beta}} \frac{1}{\sqrt{2}\pi\tau_m} 
\sqrt{I+ \sqrt{I^2+\Delta^2}} .
\label{fIcurve}
\end{equation}
QSFA does not alter the shape of the f-I curve, 
but only scales it by a factor $1/\sqrt{1+\beta}$, which 
allows us to borrow the parametric formula for the SN boundaries 
in~\cite{MPR15} (corresponding to $\beta=0$) and use it for any value of $\beta$
\footnote{
In parametric form, the SN curves ---shown in \cref{fig:Hopf_bound}--- are:
$$\bar \eta_{SN}= 
-(1+\beta) (\pi \tau_m R^*)^2 - \frac{3 \Delta^2}{4 (1+\beta) (\pi \tau_m R^*)^2},$$
 and 
$$J_{SN}= 
2(1+\beta) \pi^2 \tau_m R^*  +\frac{ \Delta^2}{2(1+\beta)   \pi^2 (\tau_mR^*)^3}.$$
}.
The phase diagrams in \cref{fig:Hopf_bound} show two Saddle-Node (SN) bifurcation curves for various 
$\beta$ values, which meet in a cusp point. Within the region 
bounded by the SN bifurcations, asynchronous, low-activity states (LAS) coexist with PS. 
In the rest of the parameter space there exists a unique fixed point that 
represents an asynchronous state.  

In the absence of adaptation, LAS and PS are both stable in the
gray-shaded region in \cref{fig:Hopf_bound}(a). 
For increasing levels of adaptation, the PS is destabilized via a Hopf bifurcation, leading to 
collective oscillations in the yellow and red shaded regions of the diagram \cref{fig:Hopf_bound}(b) 
\footnote{See \cref{sec:destroy_bistable} and Figs.~\ref{fig:bistable},\ref{fig:chaos_smallG} 
for a more detailed picture of 
this bifurcation scenario.}. 
For small $\beta$, the region where oscillations are the unique attractor (red-shaded in \cref{fig:Hopf_bound})
is restricted to a loop that pokes out of the cusp-shaped SN boundaries.
As $\beta$ is increased, the loop grows bigger and eventually unfolds almost parallel to the 
$\bar\eta$-axis, leading to a vast region of 
oscillations in parameter space, see \cref{fig:Hopf_bound}(c,d) and \cref{fig:largeG}.
Thus, sufficiently strong adaptation always leads to collective oscillations
(provided that the strength of recurrent excitation $J$ is large enough). This even occurs 
for $\bar \eta<0$, that is, in networks 
in which the majority of the neurons are quiescent in absence of recurrent excitation. 

The enhancement of collective oscillations by adaptation is greatly favored by the 
effects described in \cref{sec:micro}, concerning the distribution of the neurons' firing 
frequencies, which are also clearly reflected in the FRE~\eqref{fre3}: 
The level of adaptation $\beta$ effectively 
reduces heterogeneity $\Delta$ by a factor $1/(1+\beta)$, without altering the 
proportion of self-sustained oscillatory neurons in the population 
(by virtue of the reduction of the net input $\bar \eta$ by the same factor). This 
homogenization of the oscillators' natural frequencies promotes the emergence of 
collective synchronization~\cite{Win67,Kur84}, 
which manifests at the collective level in the form of large-scale oscillations.

Finally, we investigate in more detail the bifurcations of the FREs~(\ref{fre3},\ref{A_FRE}) for $\beta=1$,
and demonstrate that the FREs perfectly predict and replicate the collective dynamics 
of the spiking network model Eqs.~(\ref{model0},\ref{f},\ref{I}).
\cref{fig:largeG} shows a detailed picture of the phase diagram in \cref{fig:Hopf_bound}(d). 
First, we point out that the Hopf bifurcation is supercritical for positive values of $\bar \eta$, and 
becomes subcritical around $\bar \eta \approx -1$, in a generalized Hopf bifurcation (dark-red dot). This   
gives rise to a small region of bistability between the asynchronous fixed point and a limit cycle
(around the yellow star in \cref{fig:largeG}), 
which is destroyed in a Saddle-Node bifurcation of limit cycles (SNLC).
Additionally, immediately after the SNLC bifurcation crosses the lower SN bifurcation
---entering the region of coexistence between LAS and PS---, 
the stable limit cycle collides with the saddle point created in the SN bifurcation (blue dot), 
and oscillations are lost in a 
homoclinic bifurcation (blue dashed line). 
On the other hand, we find that collective oscillations also undergo period-doubling bifurcations, 
which are always present for positive $\bar \eta$. 
Inspecting the region within the period-doubling boundary more closely, reveals a
period-doubling cascade leading to macroscopic chaos. 
Collective chaos can already be found for small values of QSFA-strength $\beta$
and thus seems a generic dynamic feature of networks of QIF neurons with 
adaptation, see \cref{sec:bif_diagrams}.

%
\begin{figure}[t]
\includegraphics[width=1\columnwidth]{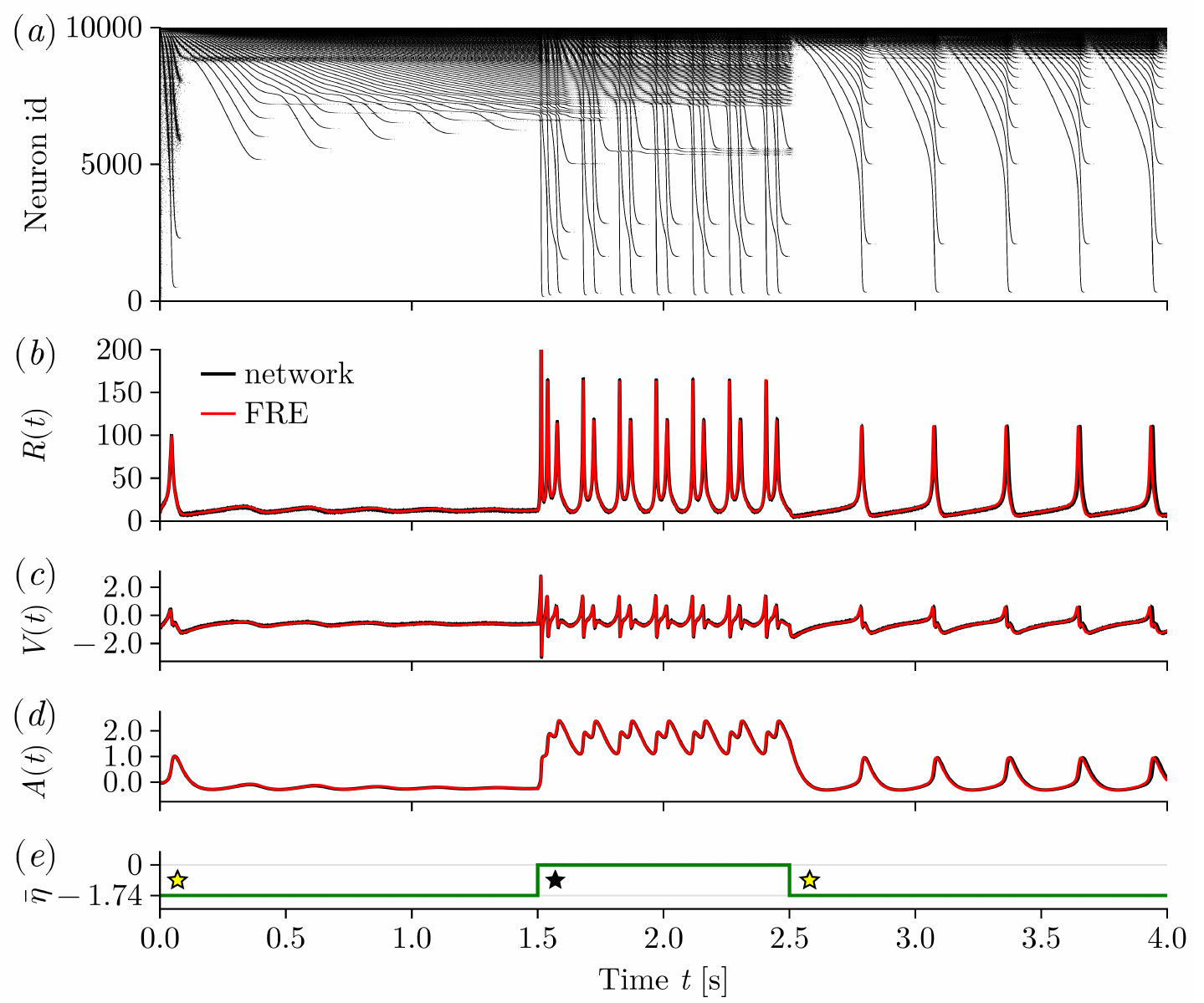}
\caption{QIF network simulations of $N=10^4$ neurons with QSFA follow the exact
FRE~(\ref{fre},\ref{Bsol},\ref{A_FRE}).
An asynchronous low-activity state coexists with network bursts (cf.~dynamics for 
$t\leq1.5$s with those for $t>2.5$s), while an increase in external input drives 
the collective dynamics into complex collective oscillations ($1.5<t\le2.5$s). 
From top to bottom: 
Raster plot of the neurons ordered according to their inputs $\eta_j$, 
population firing rate $R(t)$ (in Hz), mean voltage $V(t)$, mean adaptation $A(t)$.
Parameters: $\tau_m = 10$ms, $\tau_a=100$ms, $J=10$, $\Delta=1$
and $\bar \eta=-1.74$,  for $t\leq1.5$s and $t>2.5$s, and $\bar \eta=0$ for 
$1.5<t\le2.5$s, see yellow and black stars in \cref{fig:largeG}.
For simulation details, see \cref{sec:nums}.
}
\label{fig:net_sim}
\end{figure}

In \cref{fig:net_sim}, we compare the dynamics of the FREs~(\ref{fre},\ref{Bsol},\ref{A_FRE}),
with that of the original network model Eqs.~(\ref{model0},\ref{f},\ref{I}), 
using numerical simulations. 
We show time series of the mean field variables $R$, $V$, and 
$A$ for the two models, as well as a raster plot of the microscopic network.  
We initially set the parameters of the models in the bistable region 
of \cref{fig:largeG} ---indicated with a yellow star--- 
and select initial conditions in such a way that the 
systems converge to the asynchronous fixed point. Then, at $t=1.5$s, the input $\bar \eta$ 
instantaneously increases from $\bar \eta=0$ to $\bar \eta=1.74$, 
and the systems are placed in a region near the period-doubling bifurcation---black star in \cref{fig:largeG}. 
As the systems transition from the asynchronous regime to 
the new oscillatory state, they display identical transitory dynamics. 
Finally, at $t=2.5$s, the parameter $\bar \eta$ instantaneously returns to its initial value, 
but now the systems do not return to the fixed point, but they 
are attracted to the stable limit cycle. 
These simulation results confirm the validity of the low-dimensional 
FREs~(\ref{fre},\ref{Bsol},\ref{A_FRE}) to faithfully predict and reproduce the dynamics of the original, 
high-dimensional network model.

\begin{figure}[t]
\includegraphics[width=0.9\columnwidth]{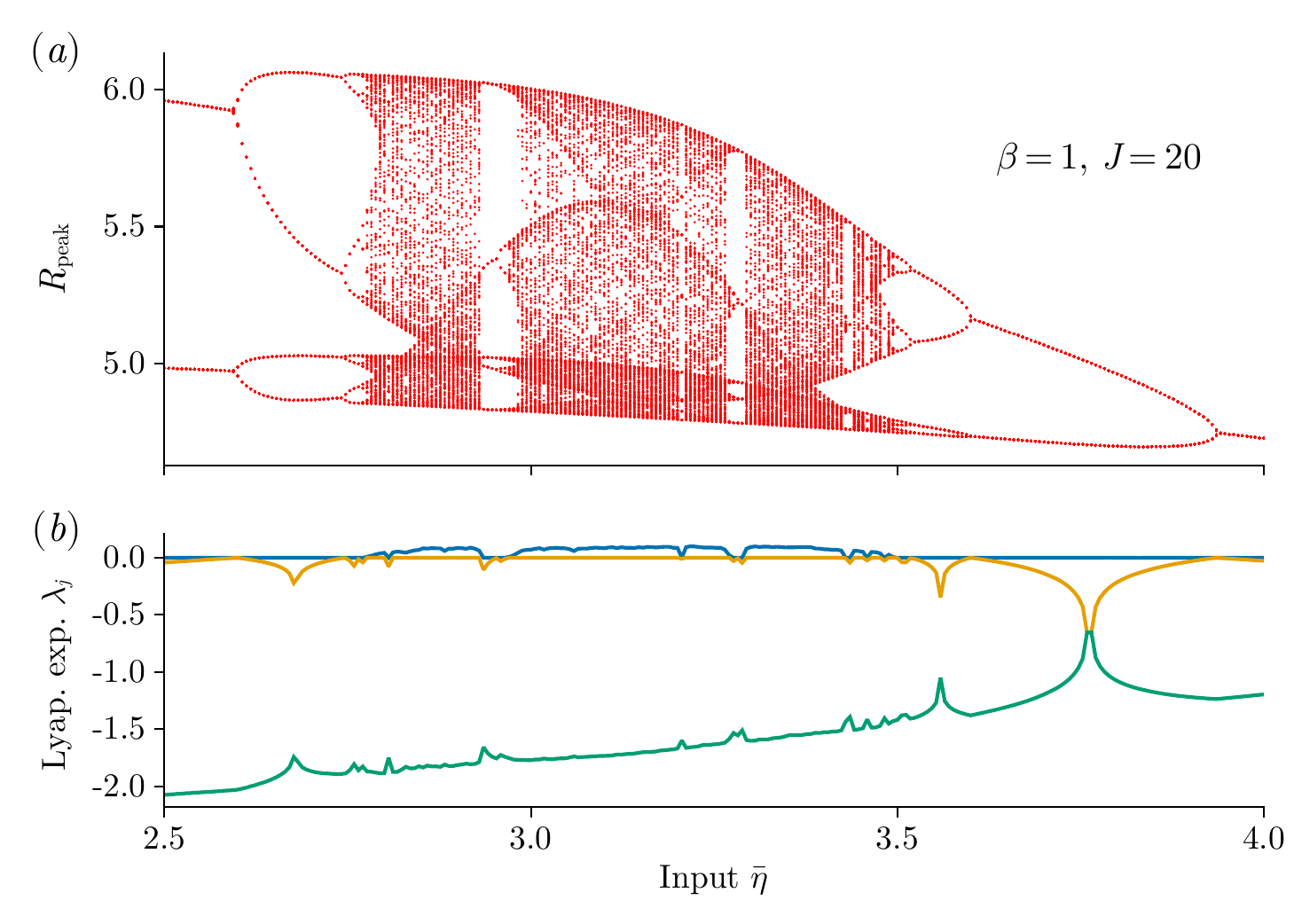}
\caption{Macroscopic chaotic behavior in the QIF model with QSFA emerges through a period-doubling cascade to chaos.
(a) Peak values $R_\text{peak}$ of the population's firing rate $R(t)$ during complex 
oscillatory activity. (b) Lyapunov exponents
computed with the FREs~(\ref{fre3},\ref{A_FRE}). A positive Lyapunov exponent 
indicates the presence of macroscopic chaos. Parameters as in \cref{fig:largeG}.
}
\label{fig:chaos}
\end{figure}
%

\section{Conclusions}
\label{sec:discussion}

Firing rate models have been exactly derived for populations of QIF neurons~\cite{MPR15,LBS13}, 
and extended to incorporate various forms of synaptic 
transmission~\cite{CB19,Lai15,RP16,PM16,PDR+19,MP20,DRM17,CKG+23,Pie24}, 
connectivity structures~\cite{Lai14,ERA+17},  
neuronal heterogeneities~\cite{PP21,DT18,GSK23}, and noise~\cite{CM24,PCP23,PP23,GPK23,GVT21,Gol21,VSG+22}. 
However, the reduction method to obtain exact FREs is limited to ensembles of one-dimensional QIF neurons. 
This restriction poses a challenge for investigating networks that exhibit important
dynamical features, such as spike-frequency adaptation (SFA).

In this work, we propose a QIF model that incorporates a quadratic SFA variable, whose evolution 
depends solely on the parameters of the QIF model and not on an individual neuron's spike train. 
This feature effectively renders the model one-dimensional, but it retains the characteristic slowing 
down of the neuron's firing frequency in response to an injected current.
Due to the quadratic dependence of the adaptation variable on the neuron's firing rate, 
the adaptation currents asymptotically match the distribution of the QIF model's input currents.
Consequently, the reduction method originally proposed in \cite{MPR15} can be applied.
The resulting exact FREs capture the neuron-specific nature of SFA---neurons with higher firing 
rates undergo greater adaptation than those with lower firing rates---,
which is reflected in the FREs as a decrease in population heterogeneity and a 
global reduction in activity.

Numerous studies have investigated the mechanisms by which SFA synchronizes neural  
firing~\cite{CEB98,VH01,FMT02,EMG01,GMG07,LAS+12,ALB+17}. 
However, to the best of our knowledge, the potential of SFA to reduce frequency heterogeneity 
within a neuronal population has not been addressed. 
In Fig.~\cref{fig:fr_distributions}, we show how this adaptation-induced homogenization markedly 
enhances the emergence of global oscillations in the network.
We further confirm that the same homogenization effect occurs in networks of heterogeneous neurons with linear SFA, suggesting that synchronization is likely to be enhanced in these networks as well. 

Additionally, we note that in mean field models of heterogeneous QIF neurons without 
neuron-specific SFA, the synchronization region is restricted to a narrower parameter space 
(cf.~Fig.~2 in \cite{FAT+23}). In these models, SFA uniformly reduces the intrinsic currents 
across neurons, rendering intrinsically spiking neurons 
quiescent under strong adaptation conditions. This suppression of firing ultimately leads to the disappearance of synchronization in regimes of strong adaptation.

In the QIF model with QSFA, we observed macroscopic chaotic behavior characterized by a period-doubling route to chaos, 
which already appears at small levels of SFA (\cref{fig:chaos_smallG}) and becomes more 
pronounced with stronger SFA (\cref{fig:chaos}). 
Interestingly, neither collective chaos nor subcritical Hopf bifurcations have been reported in firing rate models 
for QIF neurons with global SFA~\cite{GSK20,FAT+23}. 
However, a similar generalized Hopf point---separating subcritical from 
supercritical Hopf bifurcations---was identified in~\cite{GMG07}, 
along with a large region of collective oscillations for strong recurrent excitation and adaptation, 
in agreement with our findings.

Our numerical simulations of the exact FRE~(\ref{fre3},\ref{A_FRE}) closely follow those of the 
original network model Eqs.~(\ref{model0},\ref{f},\ref{I}), 
as expected, see \cref{fig:net_sim}. Still, we advise caution when interpreting results from microscopic network 
simulations due to the presence of finite-size fluctuations. 
In the QSFA model~(\ref{model0a},\ref{f}), the adaptation variable is allowed to take on negative values.
This `negative adaptation' increases the excitability of quiescent neurons by reducing the distance 
between their resting potential and the spiking threshold, 
allowing finite-size fluctuations to induce population bursts 
that would not occur in infinitely large networks, or in SFA models 
where adaptation is constrained to non-negative values

Finally, an interesting direction for future research would be to consider populations of neurons subject to stochastic inputs. 
Recent studies have investigated how different types of noise can be incorporated and analyzed within the theoretical framework of 
mean field models, such as the one investigated here, see e.g.~\cite{CM24,PCP23,PP23,GPK23,GVT21,Gol21,VSG+22,KK22,GVT23,VBM23}. 
However, it remains an open question whether some of these findings can be extended to the QIF model with QSFA proposed here.

\paragraph*{Acknowledgements.}
B.P. has received funding from the European Union’s Horizon 2020 research and innovation programme
 under the Marie Sklodowska-Curie grant agreement No 101032806. EM acknowledges support 
by the Agencia Estatal de Investigaci\'on 
under the Project No.~PID2019-109918GB-I00, and by the Generalitat de Catalunya under the 
grant 2021 SGR0 1522 646. 



\newpage
\appendix 
\setcounter{figure}{0}
\renewcommand\thefigure{S\arabic{figure}}

\section{Numerical simulation of QIF neurons with QSFA}
\label{sec:nums}
Microscopic network simulations of QIF neurons with QSFA, Eqs.~(\ref{model0},\ref{f},\ref{I}), were performed using the equivalent $\theta$-neuron formulation via $V_j = \tan(\theta_j/2)$~\cite{Erm96}:
\begin{align*}
\tau_m \dot \theta_j &= 1-\cos\theta_j + (1+\cos\theta_j) [ \eta_j - a_j + J \tau_m R(t)] , \\
\tau_a \dot a_j &= -a_j + \beta [\eta_j - a_j + J \tau_m R(t)] ,
\end{align*}
with time step $dt=10^{-3}\tau_m$, $\tau_m=10$ms and $\tau_a=100$ms.
The mean firing rate $R(t)$ was computed via the conformal mapping of the complex-valued Kuramoto order parameter $Z(t)$~\cite{MPR15,PCP23}:
\begin{equation*}
R(t) = \frac{1}{\tau_m\pi} \mathrm{Re} \left\lbrace \frac{1-Z^\ast(t)}{1+Z^\ast(t)}\right\rbrace, \quad Z(t) = \frac{1}{N} \sum_{j=1}^N e^{i\theta_j(t)};
\end{equation*}
the asterisk denotes complex conjugation.
The mean voltage was computed as $V(t) = \mathrm{Im} \lbrace [1-Z^\ast(t)]/[1+Z^\ast(t)]\rbrace$.
The mean adaptation, $\bar A(t) =\sum_{j=1}^N a_j(t)/N$, converges to $A(t)$ in the limit $N\to\infty$.

In \cref{fig:net_sim} we show numerical simulations of the FREs~(\ref{fre},\ref{Bsol},\ref{A_FRE}) 
using initial conditions $R(0)=0.01~\text{ms}^{-1}$, $V(0)=-0.2$ and $A(0)=0$ (and $\gamma=0$, $\bar c=0$).
These results are compared with numerical simulations of a network of $N=10^4$ QIF neurons with QSFA.
The voltage variables $V_j(0)$ are initially distributed according to a Lorentzian distribution centered at 
$y_0=V(0)$, and half-width at half-maximum $x_0=(\tau_m\pi)R(0)$. The adaptation variables 
follow a $\delta$-distribution, $a_j(0)=0$.

\section{Analysis of the QIF-FRE with QSFA}
\label{sec:bif_diagrams}
\subsection{SFA destabilizes persistent states}
\label{sec:destroy_bistable}
\begin{figure}[!ht]
\includegraphics[width=1\columnwidth]{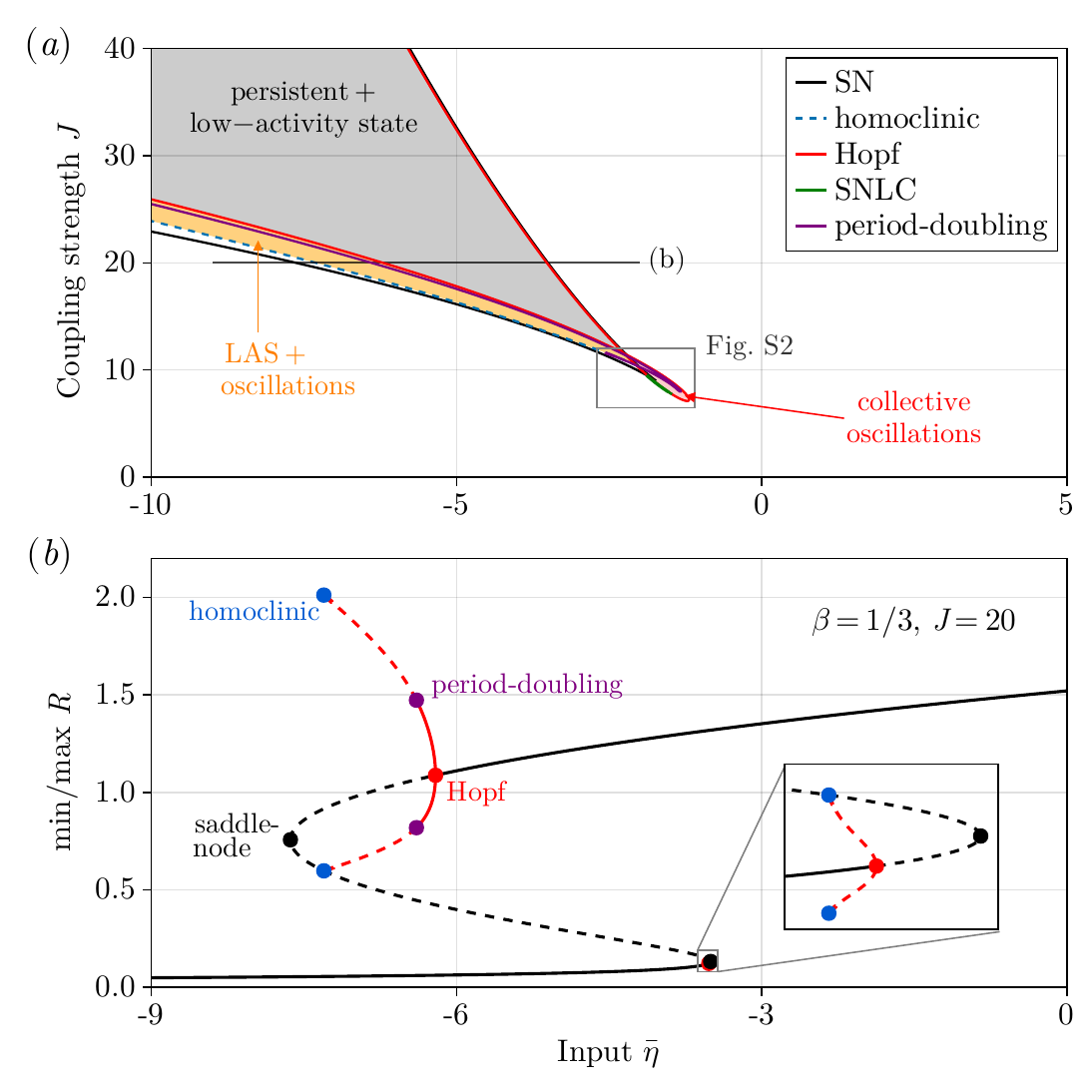}\vspace{-1em}
\caption{SFA destroys bistability. 
(a) Phase diagram in $\bar\eta-J$--plane for $\beta=1/3$. 
(b) Bifurcation diagram $R^*$ vs.\ $\bar\eta$ for $\beta=1/3, J=20$.}
\label{fig:bistable}
\end{figure}

The phase diagram for the collective dynamics of QIF neurons without QSFA (see \cref{fig:Hopf_bound}(a) with $\beta=0$) is dominated by Saddle-Node (SN) bifurcation curves that form a cusp-shaped bistability region, where two asynchronous states coexist: a low-activity state (LAS) and a high-activity, so-called persistent state (PS).
The cusp-shaped region is similar with and without QSFA by virtue of the f-I curve $\Phi$, cf.~\cref{fIcurve} and \cref{fig:Hopf_bound}.
However, already small values of QSFA induce oscillatory instabilities, where the stationary LAS and PS lose stability via Hopf bifurcations, see, e.g.,~\cref{fig:bistable} for $\beta=1/3$.
Here, the subcritical Hopf bifurcation on the lower branch occurs just before the SN point, gives rise to an unstable limit cycle solution, and therefore slightly cuts back on the bistability region.
The supercritical Hopf bifurcation on the top branch gives rise to a stable limit cycle solution and, thus, also cuts back on the bistability region of coexisting LAS and PS (the gray-shaded region in \cref{fig:bistable}a).
Moreover, the limit-cycle solution soon undergoes a period-doubling bifurcation and ends in a homoclinic bifurcation (also before the SN bifurcation, see \cref{fig:bistable}b). 
As the limit cycle has quite a constricted basin of attraction---we invite the interested reader to actually find the cycling solution for a given mean input $\bar\eta$---, the bistability between the LAS and the oscillatory solution (yellow-shaded in \cref{fig:bistable}a) de facto collapses to the LAS.
In sum, QSFA destabilizes persistent states and destroys bistability.

\subsection{Collective oscillations, network bursts, and chaos}
\label{subsec:chaos_smallG}
\begin{figure}[t]
\includegraphics[width=1\columnwidth]{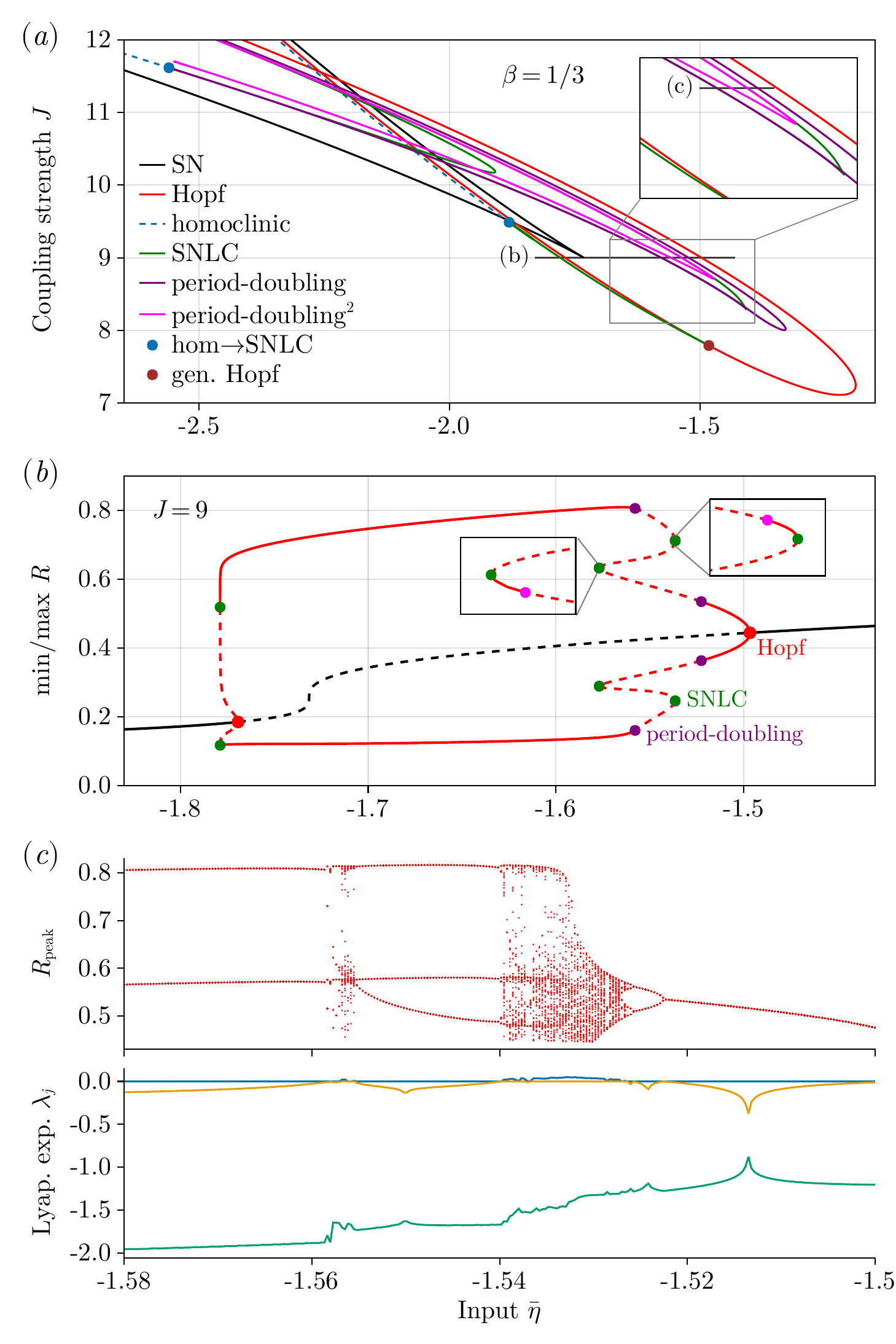}\vspace{-1em}
\caption{Collective oscillations and macroscopic chaos exist already for small QSFA. 
(a) Phase diagram for $\beta=1/3$ (zoom into the loop of \cref{fig:bistable}a) with more complicated bifurcation lines. 
(b) Bifurcation diagram $R^*$ vs.\ $\bar\eta$ for $\beta=1/3, J=9.0$.
(c) Route to chaos and first three Lyapunov exponents along the black line in the inset in (a).}
\label{fig:chaos_smallG}
\end{figure}

Collective oscillations become the dominant, and unique, attractor in the phase diagram with QSFA ($\beta>0$).
For small QSFA-strengths, $\beta=1/3$, the bifurcation scenario is somewhat intricate:
Collective oscillations are constrained to the loop that pokes out of the cusp-shaped (mostly unstable) SN boundaries (see \cref{fig:chaos_smallG} for a zoom into the loop in \cref{fig:bistable}a).
In the bifurcation diagram \cref{fig:chaos_smallG}(b), we fix the recurrent excitation at $J=9$ and decrease the mean input $\bar\eta$ from $-1.4$ to $-1.8$.
At $\bar\eta \approx -1.5$, the PS destabilizes through a supercritical Hopf bifurcation and gives rise to stable periodic oscillations. 
Subsequently, the oscillatory state undergoes a cascade of period-doubling bifurcations into macroscopic chaos around $\bar\eta\approx -1.53$ (\cref{fig:chaos_smallG}c).
Close to these parameter values, there are tiny regions where the single-periodic solution regains stability (solid green curves in the insets of \cref{fig:chaos_smallG}b), which are bounded by Saddle-Node of Limit Cycle bifurcations (SNLC, green dot) and period-doubling bifurcations (magenta dot).
For smaller $\bar\eta<-1.56$, the single-periodic solution restabilizes, though the time series of the firing rate $R(t)$ feature two peaks during each cycle (\cref{fig:chaos_smallG}c).
Around $\bar\eta\approx-1.78$, the periodic solution loses stability in a SNLC bifurcation and the unstable branch connects to the LAS in a subcritical Hopf bifurcation (red dot in \cref{fig:chaos_smallG}b), creating a small region of bistability between a limit cycle and the LAS.

\subsection{Stronger QSFA facilitates collective oscillations and chaos}
\label{subsec:strongerSFA}

As shown in \cref{fig:Hopf_bound},
the region of collective oscillations increases for larger QSFA-strengths $\beta$.
The tiny loop of dominant oscillatory collective behavior in \cref{fig:bistable} first becomes larger
and eventually completely unties as
collective oscillations expand into the $\bar\eta>0$-plane (\cref{fig:Hopf_bound}b-d).
For intermediate QSFA strengths, collective oscillations outside of the tiny loop require a substantial amount of recurrent excitation ($J\gg30$ for $\beta=1/2$, see \cref{fig:Hopf_bound}c).
The larger $\beta$, the stronger the activity-dependent self-inhibition and moderate recurrent excitation suffices to
generate oscillatory collective dynamics.  
At the same time, the intricate bifurcation structure inside the loop dissolves and more complex oscillatory behavior can safely be 
confined within a region bounded by a period-doubling bifurcation (purple curve in \cref{fig:largeG}).
Here, the macroscopic chaos emerges more clearly through a period-doubling cascade and for quite a large range of parameter values (\cref{fig:chaos}).
In sum, stronger QSFA facilitates collective oscillations and macroscopic chaos, but these generic features can also be obtained for small QSFA as seen in \cref{subsec:chaos_smallG}.

\vfill

\end{document}